\documentclass[%
 reprint,
 amsmath,amssymb,
 aps,
prb, citeautoscript,
]{revtex4-1}

\usepackage{graphicx}% Include figure files
\usepackage{dcolumn}% Align table columns on decimal point
\usepackage{bm}% bold math
\usepackage{color, soul}
\usepackage{natbib}
\usepackage[colorlinks, linkcolor=blue, citecolor=blue, urlcolor=blue]{hyperref}
\makeatletter
\newcommand*{\rom}[1]{\expandafter\@slowromancap\romannumeral #1@}
\makeatother

\begin{document}

\preprint{APS/123-QED}

\title{Machine Learning-Assisted Discovery of Carbon-Related Defect Formation Mechanisms in Diamond}

\author{Grace Huang}
\author{Christopher Fivecoat}
\author{Xun Xu}
\author{Mengen Wang}
\email{mewang@unc.edu}
\affiliation{%
Department of Physics and Astronomy, University of North Carolina at Chapel Hill}%

\begin{abstract}
Diamond is an important host material for functional defects such as solid-state spin qubits.
During the defect creation by irradiation and subsequent thermal annealing, carbon-related defects, including vacancies and interstitials, are generated and can migrate to form new defect complexes.
To investigate this kinetically driven process and reveal the formation mechanisms of the di-interstitials and di-vacancies in the neutral charge state, we fine-tuned a universal machine learning interatomic potential using a small dataset generated from density functional theory calculations.
We obtained a barrier of 1.84 eV for mono-interstitial migration and found that two initially separated mono-interstitials only need to overcome a lower barrier to form di-interstitial configurations. 
Beyond the training data, the MLIP also identified the formation pathway for an additional di-interstitial configuration and a stepwise divacancy dissociation mechanism.
These results reveal migration-assisted defect complex formation mechanisms and demonstrate how targeted MLIP training can identify reaction pathways.
\end{abstract}

\maketitle

Diamond hosts a diverse family of optically active defects with applications in quantum sensing and single-photon emission.\cite{bradac2019quantum, bassett2019quantum}
Examples include the nitrogen-vacancy, silicon-vacancy, and oxygen-vacancy centers, which have attracted interests due to their optical and spin properties.\cite{childress2013diamond,sipahigil2014indistinguishable,mukherjee2026OV}
These functional defects have relatively high formation energies and are difficult to incorporate at high concentrations under equilibrium growth conditions.\cite{weber2010quantum,filippatos2025doping}
Delta-doping or post-growth electron or ion irradiation followed by thermal annealing are commonly used to increase their concentrations.\cite{filippatos2025doping,luhmann2019coulomb}

Irradiation creates carbon vacancies and interstitials that may migrate, recombine, or be trapped by substitutional impurities to form functional defect centers.\cite{twitchen1999R2_Ci,chen2025simulating,luhmann2019coulomb}
Thermal annealing activates this migration and can convert the irradiation-induced defects into thermally stable functional centers.\cite{twitchen1999R2_Ci,luhmann2019coulomb}
First-principles computation can be used to understand the thermodynamic stability of defects at various charge states.
For example, the NV center was found to have a high formation energy and can form through the capture of a mobile carbon vacancy by a substitutional N; competing reactions such as vacancy aggregation can affect NV creation or charge-state stability.\cite{weber2010quantum, chen2025simulating,deak2014formation}

Beyond impurity-related centers, intrinsic defects including mono-vacancies, divacancies, mono-interstitials, and di-interstitials are also important in diamond processing.\cite{cherati2026monodi,slepetz2013divacancies}
Carbon interstitials can be irradiation-induced, and EPR measurements have identified the R2 center as the neutral split interstitial configuration.\cite{twitchen1999R2_Ci}
The di-interstitial configurations are also observed experimentally. However, the origin of these defect centers remains largely unexplored.\cite{twitchen1999optical_diinter, twitchen1999R2_Ci}
A recent first-principles study reported high formation energies for carbon interstitial and di-interstitial configurations in diamond, and a strong driving force for aggregation.\cite{cherati2026monodi}
It is important to understand the mechanisms of mono-interstitials aggregation and the corresponding activation barriers to form defect complexes.

Vacancy defects play an important role in diamond processing and color-center formation.\cite{weber2010quantum,luhmann2019coulomb}
Migrating monovacancies can be captured by substitutional impurities or combine to form divacancies and larger vacancy clusters.\cite{chen2025simulating,slepetz2013divacancies,luhmann2019coulomb,deak2014formation}
Vacancy aggregation reduces the population of monovacancies available to form impurity-vacancy color centers, while the resulting clusters can act as charge traps and sources of spin noise.\cite{luhmann2019coulomb}

To reveal the formation mechanisms of the defect complexes, it is important to go beyond the thermodynamic formation energies and identify the possible reaction pathways for the formation and dissociation, which informs the critical annealing temperature to facilitate the formation or annihilation of such complexes.
Nudged elastic band calculations are widely used to determine minimum-energy pathways and activation barriers.\cite{henkelman2000tangent,henkelman2000climbing}
Because several intermediate images must be optimized simultaneously, an NEB calculation is typically several times more expensive than a single structural relaxation.
To efficiently screen and identify reaction pathways, we combine machine learning interatomic potential (MLIP) with density functional theory (DFT) calculations to explore the migration of mono-vacancy and mono-interstitial, and the mechanism of formation and dissociation of di-vacancy and di-interstitials.

Universal MLIPs provide a starting point to explore migration pathways. The pretraining datasets of many universal MLIPs are dominated by equilibrium and near-equilibrium structures, which can lead to systematically underestimated defect energies and migration barriers.\cite{deng2025systematic}
We therefore generated representative training configurations from relaxations, NEB calculations, and ab-initio molecular dynamics simulations (AIMD) of carbon-related defects using a limited number of images and ionic steps.
The universal MLIP is fine-tuned and validated to reproduce the migration pathways and barriers of the defect complexes. 
We first discuss the migration of the carbon mono-interstitials and the formation mechanism of two di-interstitial configurations.
The transferability of the fine-tuned MLIP is further evaluated by applying it to predict the migration pathway and activation energy barrier for an untrained di-interstitial configuration and the dissociation mechanism of the di-vacancy.
Further DFT calculations provide the final migration mechanisms and are used to assess the MLIP in extending pathway exploration to related defect configurations.

DFT calculations are performed using the Vienna Ab initio Simulation Package (VASP) \cite{kresse1996vasp1, kresse1996vasp2}.
The Perdew-Burke-Ernzerhof (PBE) functional\cite{Perdew_pbe1996} and a plane-wave energy cutoff of 400 eV are used.
The Brillouin zone was sampled using a $1 \times 1 \times 1$ $\Gamma$-centered k-point mesh.
The computed lattice constant of diamond is 3.568$\mathrm{\AA}$, in reasonable agreement with the experimental value of 3.567$\mathrm{\AA}$\cite{yamanaka_lattice1994}.
The stability of all defects was evaluated using defect formation energy diagrams.\cite{northrup_formation_energy1997,freysoldt2014first}
We calculated the formation energies $E^\mathrm{f}$ for a carbon-related defect with a charge state $q$ using 
\begin{equation}
    E^\mathrm{f}[X^q] = E_\mathrm{tot}[X^q] - E_\mathrm{bulk} - n_\mathrm{C}\mu_\mathrm{C} +qE_\mathrm{F} + E_\mathrm{corr},
\end{equation}
where $E_\mathrm{tot}[X^q]$ and $E_\mathrm{bulk}$ are the total energies of the supercell with a defect and the perfect lattice. 
$n_\mathrm{C}$ is the number of carbon atoms added or removed.
$\mu_\mathrm{C}$ is the chemical potential of carbon calculated as the energy per atom of the bulk, and $E_\mathrm{F}$ is the Fermi level.
$E_\mathrm{corr}$ is the correction term for the charged defects \cite{freysoldt_E_corr2018}.

\begin{figure}
\includegraphics[width=85mm]{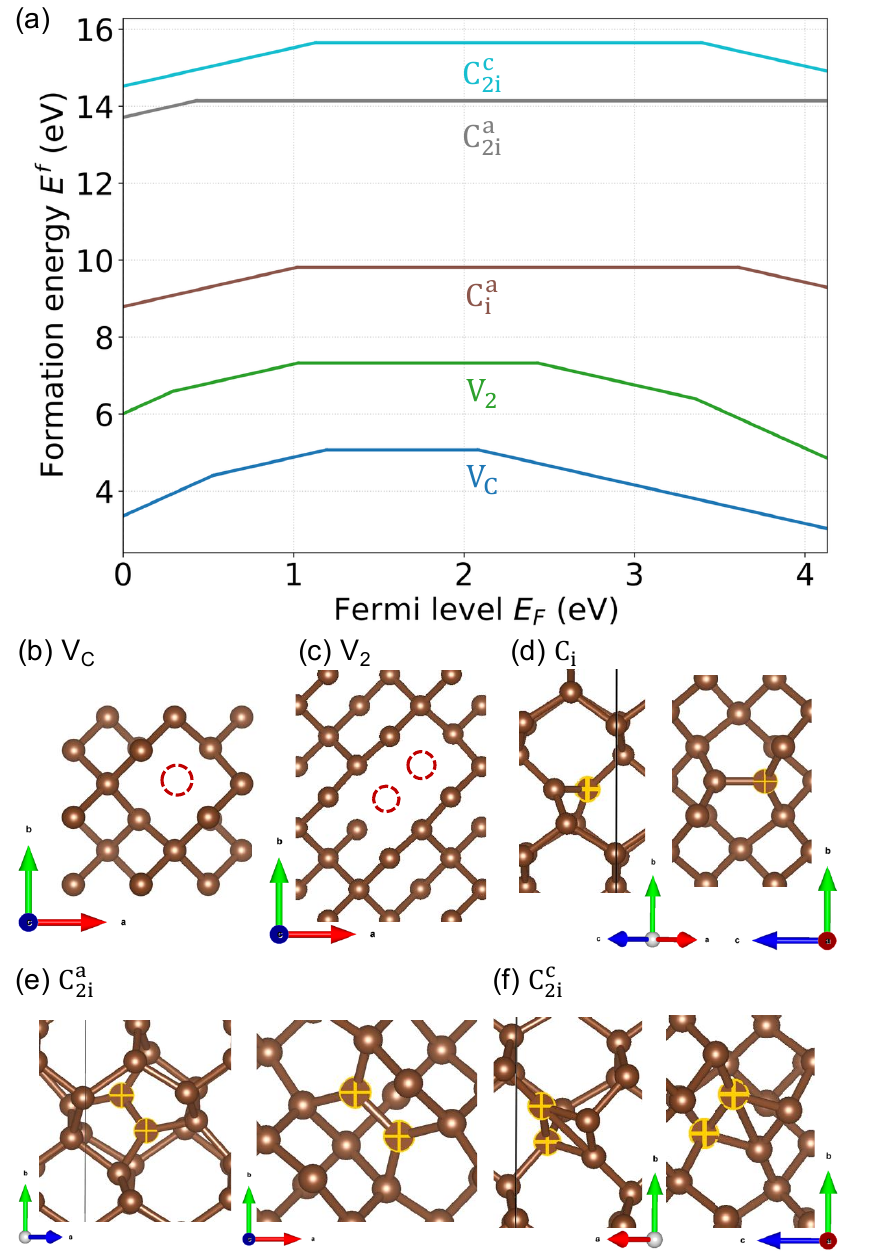}
\caption{\label{fig:energy_diagram} Defect formation energy diagram for five defects (a) and structures of  (b) V$_\mathrm{C}$, (c) V$_2$, (d) C$_\mathrm{i}$, (e) C$_\mathrm{2i}^\mathrm{a}$, and (f) C$_\mathrm{2i}^\mathrm{c}$.
}
\end{figure}

Figure \ref{fig:energy_diagram} (a) shows the formation energy diagram of the five carbon-related intrinsic defects.
We note that accurate charge transition levels require a hybrid functional such as HSE06.
Here we employ PBE results to more efficiently sample training data along the migration pathway, and we also found PBE to be useful for comparing relative defect stabilities and identifying stable charge states.
V$_\mathrm{C}$ [Fig.\ref{fig:energy_diagram} (b)] has the lowest formation energy. 
The divacancy (V$_2$) [Fig.\ref{fig:energy_diagram} (c)] is found to have a higher formation energy due to the extra energy cost to break two more C bonds as compared to the single vacancy. 
When the fermi level is high in the bandgap, V$_\mathrm{C}$ prefers the q=$-1$ charge state, while V$_2$ can be stable in q=$-1$ and q=$-2$, corresponding to the conditions of $n$-type doped diamond by nitrogen or phosphorus.\cite{rouzbahani2024ndoping}
The neutral charge state of V$_\mathrm{C}$ and V$_2$ can be stabilized when the Fermi level is pinned lower in the bandgap, corresponding to more $p$-type conditions enabled by trivalent doping such as B and Al.\cite{filippatos2025doping}

Interstitials have higher formation energies as compared to vacancies due to the lattice distortion needed to insert extra carbon atoms into the diamond lattice.
Following the prior work of systematic investigation of carbon interstitials in diamond\cite{cherati2026monodi}, we built the mono-interstitial and di-interstitial structures and found consistent structures.
The formation energies are consistently lower than HSE06 results due to the change of functional. However, we note that both functionals identified the neutral charge state of interstitials as stable under a wide range of the band gap, which is the focus of energetic discussions relevant to their formation mechanisms. 
The ground state structure of the mono-interstitial (C$_\mathrm{i}$) is a split interstitial configuration with $D_{2d}$ symmetry [Fig. \ref{fig:energy_diagram} (d)], where the original carbon atom from the perfect lattice shares the same site with the interstitial carbon atom along the [001] direction.
The distance between the two C atoms is 1.28 $\mathrm{\AA}$.
The C$_\mathrm{2i}^\mathrm{a}$ [Fig. \ref{fig:energy_diagram} (e)] structure is formed by intercalating two carbon interstitial atoms inside the same carbon ring, and each carbon interstitial atom forms two bonds with the carbon atoms with a bond length of 1.35 $\mathrm{\AA}$. 
We found that C$_\mathrm{2i}^\mathrm{c}$ [Fig. \ref{fig:energy_diagram} (f)] has a triplet ground state, consistent with the HSE06 result.\cite{cherati2026monodi}
The structure of C$_\mathrm{2i}^\mathrm{c}$ is built by two C$_\mathrm{i}$ on neighboring carbon atoms, forming a split interstitial configuration.
C$_\mathrm{2i}^\mathrm{c}$ is 1.51 eV higher in energy than C$_\mathrm{2i}^\mathrm{a}$.

\begin{figure*}
\includegraphics[width=180 mm]{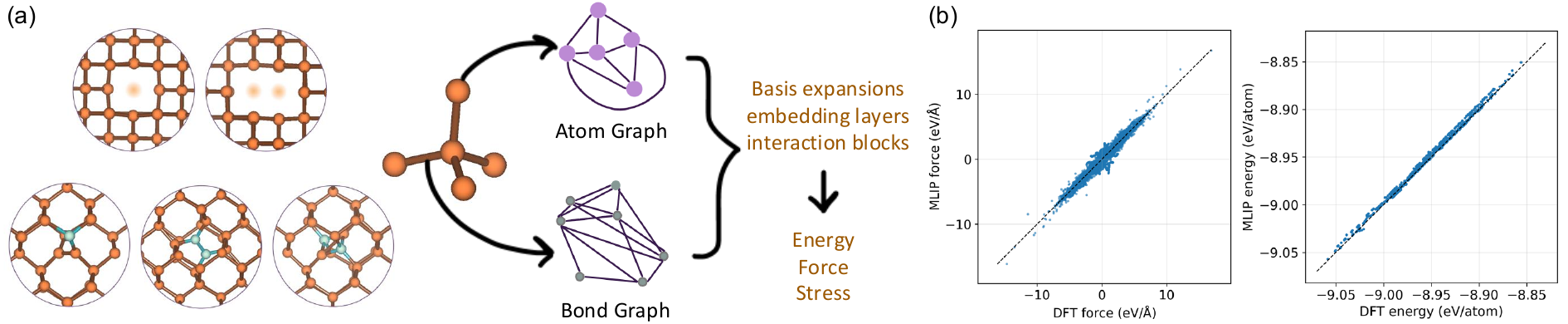}
\caption{\label{fig:overview} Overview of MLIP training and validation. (a) Illustration of the data and model used for MLIP training. (b) The parity plots comparing the MLIP-predicted energy per atom and force calculated with the DFT reference values of the independent testing dataset.
}
\end{figure*}

To better understand the stability and formation mechanisms of the intrinsic carbon-related defects in diamond, we performed nudged elastic band (NEB) calculations followed by climbing-image nudged elastic band (CI-NEB) calculations to understand the migration pathways and the corresponding activation energy barriers.\cite{henkelman2000climbing,henkelman2000tangent}  
The data generated from relaxations of the initial state and final state structures, NEB calculations, and ab initio molecular dynamics simulations (AIMD) of V$_\mathrm{C}$, V$_\mathrm{2}$, C$_\mathrm{i}$, C$_\mathrm{2i}^\mathrm{a}$, and C$_\mathrm{2i}^\mathrm{c}$ are used to train an MLIP.
To reduce the correlation between the configurations, the relaxation data were sampled every two ionic steps starting from the 5th ionic step, retaining 269 defect configurations. The NEB trajectories were sampled every 3 steps over the first 100 steps, obtaining 384 configurations. 
We note that only a small number of NEB images representing a unique migration pathway are needed in the training data.
The AIMD trajectories were sampled every 6 steps, contributing 335 configurations.
Together with a reference structure for the perfect diamond supercell, the final dataset contained 989 configurations spanning relaxation pathways, transition state structures, and finite-temperature environments (Table S1).

\begin{figure}
\includegraphics[width=90mm]{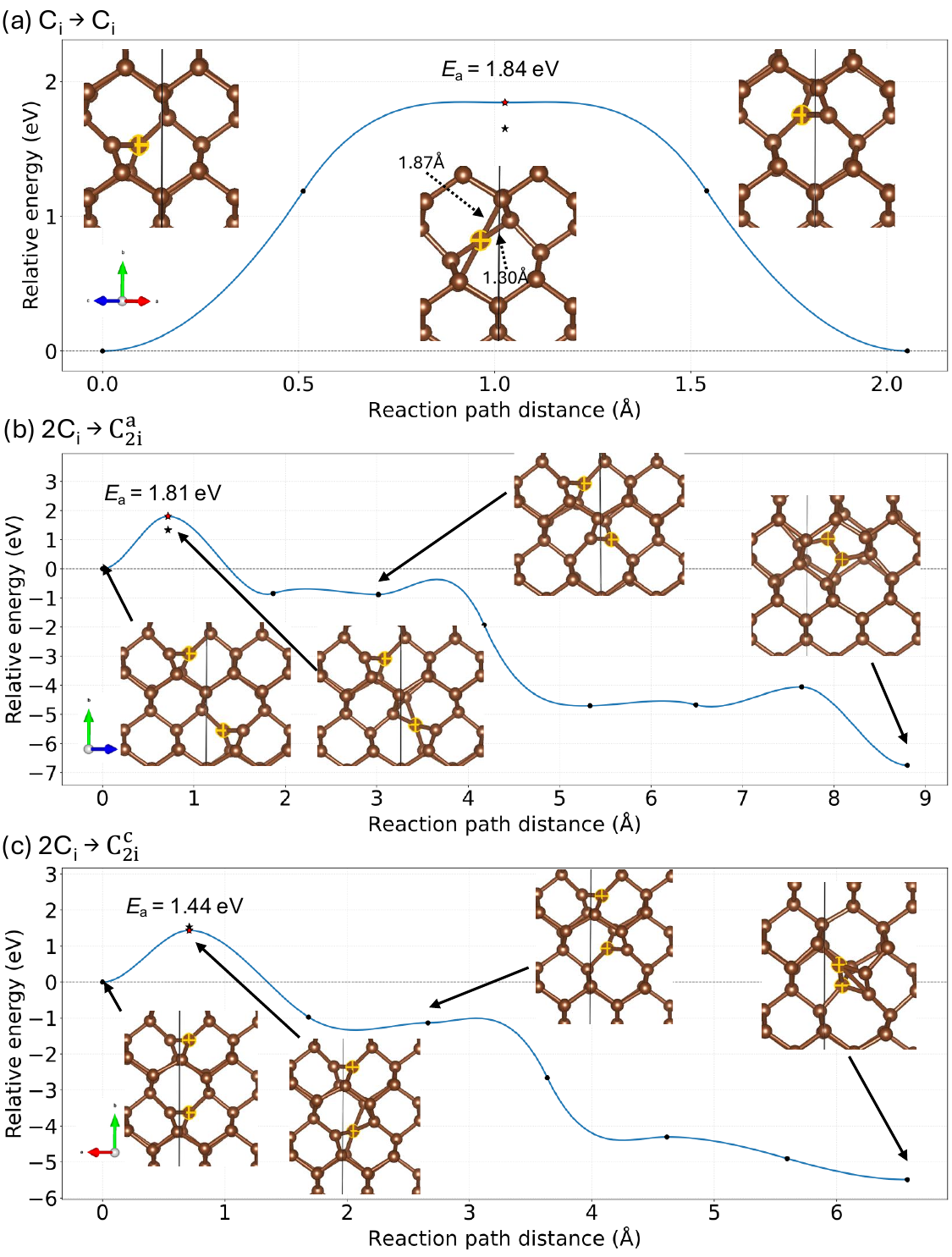}
\caption{\label{fig:migration_interstitials} The migration pathways, activation energy barriers ($E_\mathrm{a}$), and the corresponding initial state, transition state, and final state structures of (a) the migration of C$_\mathrm{i}$ and the formation of (b) C$_\mathrm{2i}^\mathrm{a}$ and (c) C$_\mathrm{2i}^\mathrm{c}$ from two C$_\mathrm{i}$ from CI-NEB calculations. The black star represents the corresponding migration barriers from CI-NEB calculations using MLIP.
}
\end{figure}

\begin{figure*}
\includegraphics[width=180mm]{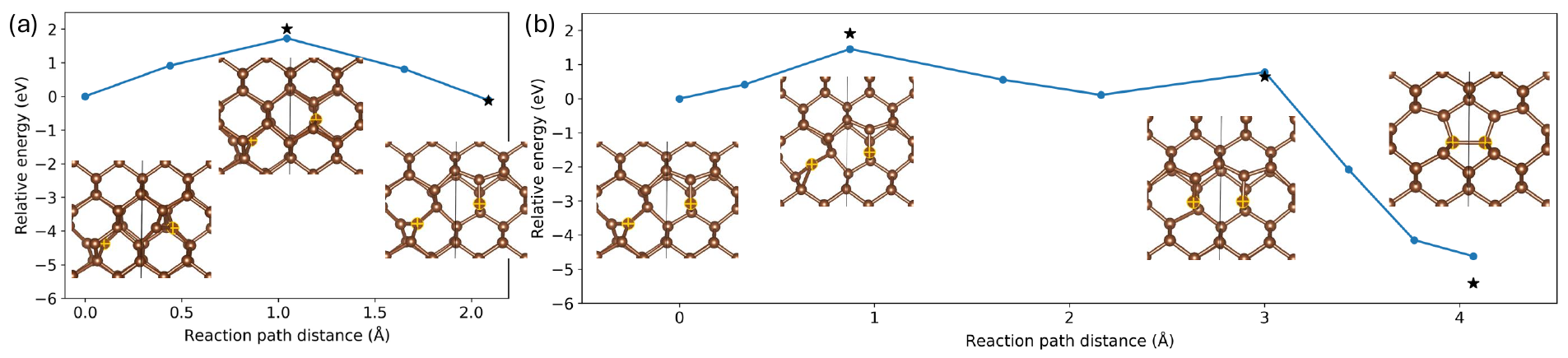}
\caption{\label{fig:2cib} The two-stage MLIP-predicted formation pathway of C$_\mathrm{2i}^\mathrm{b}$ from two nearby C$_\mathrm{i}$ defects. (a) The two C$_\mathrm{i}$ defects approach to form a locally stable intermediate state. (b) The intermediate transforms into the final C$_\mathrm{2i}^\mathrm{b}$ configuration. Blue lines and circles show the MLIP energy profiles, and black stars show DFT single-point energies calculated for MLIP-generated structures.}

\end{figure*}

\begin{figure}
\includegraphics[width=87mm]{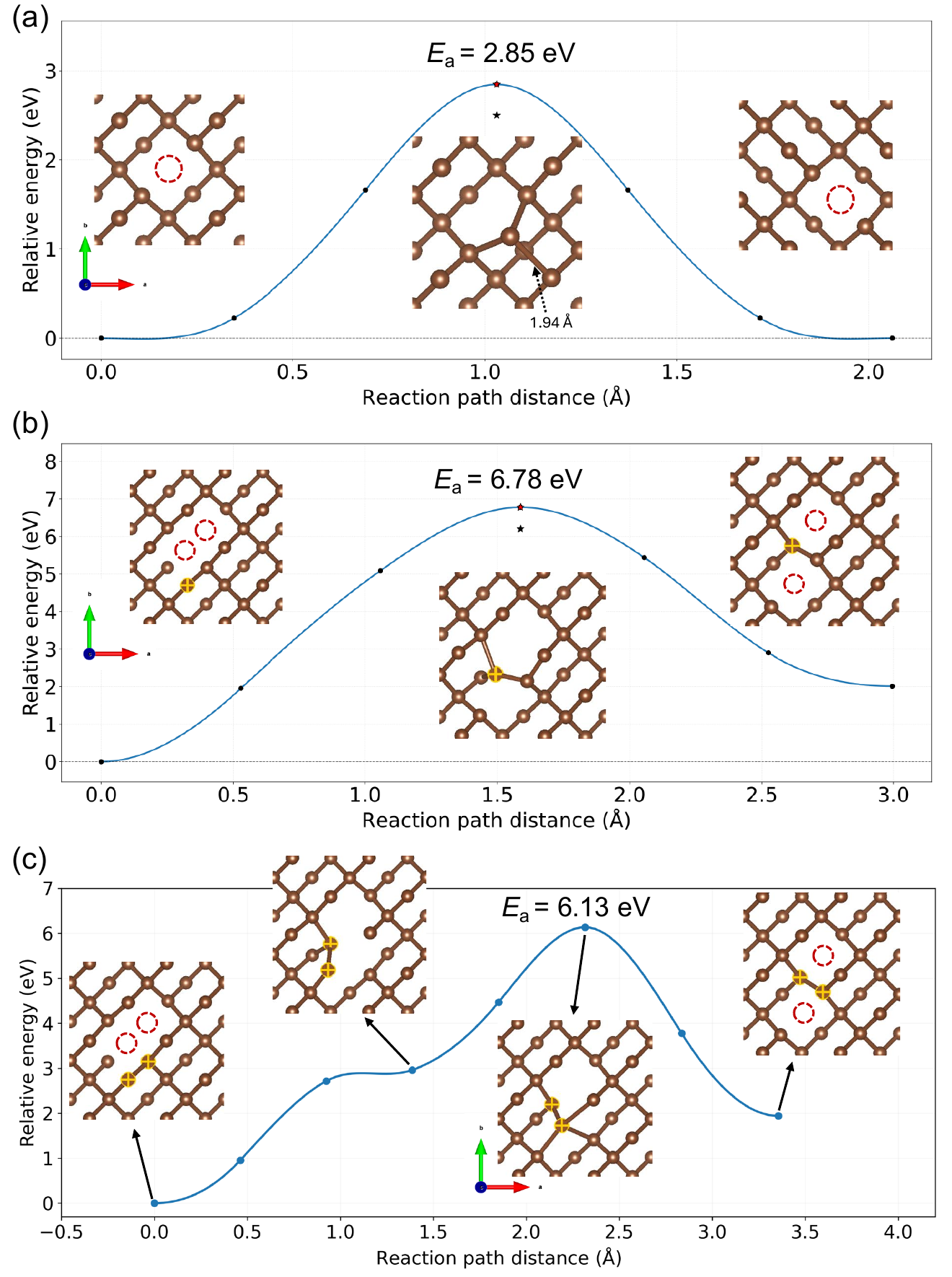}
\caption{\label{fig:migration_vacancies} The migration pathways, activation energy barriers ($E_\mathrm{a}$), and the corresponding initial state, transition state, and final state structures of (a) V$_\mathrm{C}$ migration, (b) V$_2$ dissociation into two neighboring V$_\mathrm{C}$, and (c) the step-wise dissociation pathway of V$_2$.
}
\end{figure}

Figure \ref{fig:overview} (a) shows an overview of the machine learning-assisted defect migration pathway exploration.
A pre-trained Crystal Hamiltonian Graph Neural Network (CHGNet) model\cite{deng_chgnet} was fine-tuned with training targets of energy, force, and stress, with loss ratios of 5, 1, and 0.1. 
The training employed the Huber loss function, the Adam optimizer, and a learning rate of 10$^{-4}$ with a cosine learning rate decay fraction of 10$^{-2}$. 
The model is trained on 90\% of the 989 data structures for 300 epochs with no weight decay. The remaining 10\% of structures are used for validation. 
To evaluate the predictive performance of the fine-tuned MLIP, energy and force predictions are compared with the corresponding DFT values for an independent dataset containing structures not included in the 989 training dataset.
The testing dataset includes 3755 structures, including structural relaxations, NEB calculations, and AIMD trajectories (Table S2).
While the NEB configurations used for fine-tuning were sampled from the first 100 ionic steps, the testing dataset includes configurations sampled throughout the available trajectories beyond the 100 steps. 
It also includes additional NEB images for V$_2$ dissociation, which are not included in training.
The energy and force parity plots [Figs. \ref{fig:overview} (b)] illustrate good agreement between model predictions and the DFT values.
The fine-tuned MLIP was used to perform CI-NEB calculations to identify reaction pathways, followed by additional calculations from DFT CI-NEB for the final activation barriers.
The reaction pathways, activation barriers, and the intermediate state structures from DFT are shown in Figs. \ref{fig:migration_interstitials} and \ref{fig:migration_vacancies}, as compared to the results from MLIP (Figures S1 and S2).
MLIP slightly underestimates the activation barrier. We note that the largest error (0.58 eV) comes from V$_\mathrm{2}$ dissociation due to the lack of training data from V$_\mathrm{2}$ NEB pathways.
The MLIP and DFT results show consistent structures along the migration pathway, supporting the use of the fine-tuned MLIP to identify reaction pathways.

We start with the discussion of the formation and migration mechanisms of the interstitials.
The formation energy of the mono-interstitial is 9.81 eV in the neutral charge state, indicating that it is unlikely to form under thermodynamic equilibrium conditions.
Experimentally, electron irradiation is used to create carbon vacancies and place the carbon atom into an interstitial position.\cite{twitchen1999R2_Ci}
Figure \ref{fig:migration_interstitials} (a) shows the migration pathway of the mono-interstitial (C$_\mathrm{i}$).
The final state is an equivalent configuration to the initial state where the split interstitial configuration forms at a neighboring carbon site.
In the transition state structure, the carbon interstitial forms four C-C bonds, with a bond length of 1.87 or 1.30 $\mathrm{\AA}$. 
The corresponding activation energy barrier is 1.84 eV, which is consistent with the experimental value of 1.56-1.74 eV.\cite{allers1998Ci_annealing, twitchen1999R2_Ci} 
Our value is also close to the reported barrier of 1.82 eV from NVT AIMD simulations.\cite{chen2025simulating}
%The migration barrier from MLIP is calculated by full relaxation of the initial state and final state structure followed by CINEB calculations, yielding an $E_\mathrm{a}$ of 1.6 eV. 
The rate of migration can be calculated as $\Gamma=\Gamma_0e^{-E_\mathrm{a}/k_\mathrm{B}T}$, where $E_\mathrm{a}$ is the migration barrier, $k_\mathrm{B}$ is the Boltzmann constant, $T$ is the annealing temperature, and $\Gamma_0=10^{13}$ s$^{-1}$ is the pre-exponential factor.\cite{vineyard1957frequency,wang2020mechanism}
The temperature that enables the defect to become mobile (annealing temperature) can be estimated by using $\Gamma=1$ s$^{-1}$.\cite{wang2022ga2o3_diffusion}
Applying this method, the annealing temperature for C$_\mathrm{i}$ is 713 K, which is consistent with the experimental thermal annealing temperature of $\sim$700 K.\cite{twitchen1999R2_Ci}

The defect formation energies of C$_\mathrm{2i}^\mathrm{a}$ and C$_\mathrm{2i}^\mathrm{c}$ are even higher than the mono-interstitial, indicating they are harder to form.
However, we note that their binding energies, calculated with $E_\mathrm{b}=2E(C_\mathrm{i})-E(C_\mathrm{2i})$, are 5.49 eV for C$_\mathrm{2i}^\mathrm{a}$ and 3.98 eV for C$_\mathrm{2i}^\mathrm{c}$, indicating that the two mono-interstitials prefer to bind together into a di-interstitial configuration.\cite{cherati2026monodi}
To understand the formation mechanisms of C$_\mathrm{2i}^\mathrm{a}$ and C$_\mathrm{2i}^\mathrm{c}$, we calculated the migration pathways and the corresponding activation energy barriers for two initially separated C$_\mathrm{i}$ defects combining into a di-interstitial configuration.

Figure \ref{fig:migration_interstitials} (b) shows the formation of the C$_\mathrm{2i}^\mathrm{a}$ defects from two carbon interstitials with a distance of 3.91 $\mathrm{\AA}$. 
The activation energy is 1.81 eV, which is very close to the migration barrier of the C$_\mathrm{i}$.
The transition state structure also closely resembles that of C$_\mathrm{i}$ where the carbon atom forms four C-C bonds with distances ranging from 1.36 to 1.63 $\mathrm{\AA}$. 
We also found a stable intermediate structure that is 0.89 eV lower in energy than the initial state.
This intermediate state consists of two split interstitials with a distance of 2.86 $\mathrm{\AA}$.
Starting from this intermediate state, the formation of C$_\mathrm{2i}^\mathrm{a}$ is almost barrierless.

Figure \ref{fig:migration_interstitials} (c) shows the formation of C$_\mathrm{2i}^\mathrm{c}$ from two carbon interstitials with a distance of 3.69 $\mathrm{\AA}$. 
The activation energy barrier is 1.43 eV, which is slightly lower than C$_\mathrm{2i}^\mathrm{a}$ partially due to the closer C$_\mathrm{i}$-C$_\mathrm{i}$ distance.
We also found a stable intermediate state consisting of two split interstitials with a distance of 2.69 $\mathrm{\AA}$, which is 1.14 eV lower than the initial state structure.
The formation of C$_\mathrm{2i}^\mathrm{c}$ from this intermediate state is also barrierless, and the formation of the di-interstitial configuration is mainly driven by the large binding energies.
As C$_\mathrm{2i}^\mathrm{a}$ has a lower formation energy than C$_\mathrm{2i}^\mathrm{c}$, the energy gain during the reaction energy of the C$_\mathrm{i}$+C$_\mathrm{i}$ $\to$ C$_\mathrm{2i}^\mathrm{a}$ process is 1.26 eV higher than C$_\mathrm{i}$+C$_\mathrm{i}$ $\to$ C$_\mathrm{2i}^\mathrm{c}$.

The annealing temperatures calculated from the barriers of 1.81 eV and 1.44 eV are 702 K for C$_\mathrm{2i}^\mathrm{a}$ and 554 K for C$_\mathrm{2i}^\mathrm{c}$. 
We expect that, once the thermally activated diffusion of the mono-interstitial brings two C$_\mathrm{i}$ defects to a critical distance of $\sim$ 3.6 to 4 $\mathrm{\AA}$, the formation of the di-interstitials becomes kinetically favorable due to the smaller activation barriers.
The experimentally observed R1 EPR center has been associated with the C$_\mathrm{2i}^\mathrm{c}$ defect despite its relative high formation energy.\cite{twitchen1996di_inter,twitchen1999optical_diinter}
We also note that once C$_\mathrm{2i}^\mathrm{a}$ and C$_\mathrm{2i}^\mathrm{c}$ form due to the mono-interstitial migration, the dissociation requires substantially higher temperatures.
The dissociation activation barrier for C$_\mathrm{2i}^\mathrm{c}$ is 6.92 eV, corresponding to an activation temperature of 2683 K, which is within the range of high-pressure high-temperature (HPHT) annealing conditions.\cite{zhang2024hpht}
The dissociation barrier for C$_\mathrm{2i}^\mathrm{a}$ is 8.55 eV with an even higher activation temperature of 3315 K.

Besides C$_\mathrm{2i}^\mathrm{a}$ and C$_\mathrm{2i}^\mathrm{c}$, C$_\mathrm{2i}^\mathrm{b}$ is also likely to form due to the slightly lower formation energy as compared to C$_\mathrm{2i}^\mathrm{c}$.
A recent theoretical work has attributed the TR12 center to the C$_\mathrm{2i}^\mathrm{b}$ defect.\cite{cherati2026monodi}
The configuration of C$_\mathrm{2i}^\mathrm{b}$ is built by two carbon interstitials occupying two next-nearest-neighbor sites.
To identify the formation mechanism of C$_\mathrm{2i}^\mathrm{b}$, we directly applied the fine-tuned MLIP with no further training.

Although C$_\mathrm{2i}^\mathrm{b}$ was not included in the training data, the fine-tuned MLIP produced a continuous two-stage formation pathway. 
Figure \ref{fig:2cib}(a) shows the first stage, where two separated C$_\mathrm{i}$ defects approach and overcome a barrier of 1.73 eV to form a locally stable intermediate state. 
Figure \ref{fig:2cib}(b) shows the second stage, where the intermediate state transforms into the final C$_\mathrm{2i}^\mathrm{b}$ configuration with a barrier of 1.45 eV. 
To examine the energetics of the MLIP-predicted pathway, we performed DFT single-point calculations for selected transition-state, intermediate, and final-state structures generated by the MLIP. 
The DFT single-point energies are close to the corresponding MLIP values at these selected configurations, yielding a barrier of 2.01 eV.
This comparison tests the energetic consistency of representative structures along the MLIP pathway.

As vacancies and interstitials are generated together as Frenkel pairs under irradiation, we next briefly discuss this complementary vacancy-related kinetic process. 
Mobile V$_\mathrm{C}$ is an important process associated with functional defect complex formation such as the nitrogen vacancy (NV) center or the silicon vacancy (SiV) center.
Compared with the interstitial, the monovacancy requires a higher temperature to be mobile. 
The migration barrier is 2.85 eV [Fig. \ref{fig:migration_vacancies} (a)], corresponding to an annealing temperature of 1105 K, close to the typical standard annealing conditions to mobilize these vacancies toward substitutional nitrogen defects to form NV \cite{rose2018SiV, meijer2005Nannealing,chu2014coherent}. 

The binding energy for V$_\mathrm{2}$ is 2.82 eV, indicating that forming the defect complex is energetically favorable. 
To study the migration energetics of the dissociation of V$_\mathrm{2}$, or the formation of V$_\mathrm{2}$ from two neighboring V$_\mathrm{C}$, we explored two possible migration pathways.
Figure \ref{fig:migration_vacancies} (b) illustrates the migration pathway and barrier of a direct dissociation from V$_\mathrm{2}$ to two V$_\mathrm{C}$. 
In the transition state, the carbon atom associated with the vacancy migration forms a shorter bond of 1.43 $\mathrm{\AA}$ and two longer bonds of 1.81 $\mathrm{\AA}$.
Due to the bond breaking associated with the vacancy migration, the activation energy barrier for V$_\mathrm{2}$ dissociation is 6.78 eV and the reaction energy is 2.01 eV.
Even with no training data from the V$_\mathrm{2}$ dissociation pathway, the MLIP we trained still provided a reasonable activation barrier of 6.20 eV.
We expect the 0.58 eV underestimate in the activation barrier can be reduced by fine-tuning with more vacancy-related migration data.\cite{deng2025systematic}

We applied the fine-tuned MLIP to identify an alternative pathway of V$_2$ dissociation with an activation barrier of 4.89 eV.
The DFT calculation following the MLIP pathway confirms that this pathway has a lower barrier (6.13 eV) [Figure \ref{fig:migration_vacancies} (c)] than the direct dissociation pathway, for which the barrier is 6.78 eV [Fig. \ref{fig:migration_vacancies}(b)].
Although MLIP underestimates the stepwise dissociation barrier by 1.24 eV, it identifies the intermediate configuration and reproduces the lower barrier of this pathway relative to direct dissociation.
The origin of a lower activation energy barrier of the pathway discovered by MLIP is the step-wise migration mechanism.
The migration of a vacancy is intrinsically enabled by migrating atom(s) associated with the vacancy.
Figure \ref{fig:migration_vacancies} (b) shows how structures change along the direct migration pathway, where the highlighted atom migrates directly from the initial state to the final state, yielding a barrier of 6.78 eV.
In the step-wise migration pathway shown in Fig. \ref{fig:migration_vacancies} (c), there are two carbon atoms associated with the migration, facilitating the formation of an intermediate metastable state and a more energetically favorable transition state.
The step-wise formation mechanism of the divacancy is also consistent with the prior DFT results.\cite{slepetz2013divacancies}
The intermediate and transition state structures of the step-wise migration pathway resembles more closely the single-vacancy migration, which can help explain MLIP trianed with V$_\mathrm{C}$ migration data can identify reasonable reaction pathways for V$_2$ dissociation.

In conclusion, we combined DFT calculations with a fine-tuned MLIP to investigate the migration, formation, and dissociation mechanisms of intrinsic carbon defects in diamond. 
The mono-interstitial has a migration barrier of 1.84 eV.
Once carbon mono-interstitials become mobile during annealing, two C$_\mathrm{i}$ brought into nearby configurations overcome lower barriers of 1.81 eV and 1.44 eV to form stable di-interstitial configurations C$_\mathrm{2i}^\mathrm{a}$ and C$_\mathrm{2i}^\mathrm{c}$.
These two di-interstitial configurations also have high binding energies of 5.49 eV and 3.98 eV, supporting a migration-assisted formation mechanism.
The MLIP also identified a two-stage formation pathway for C$_\mathrm{2i}^\mathrm{b}$, which is not included in the training data. 
Two C$_\mathrm{i}$ need to overcome a barrier of 1.73 eV form C$_\mathrm{2i}^\mathrm{b}$.
The mono-vacancy has a migration barrier of 2.85 eV.
Using MLIP, we identified an alternative step-wise dissociation pathway involving two carbon atoms facilitating a lower activation energy. 
DFT calculations confirm that the step-wise dissociation pathway has a barrier of 6.13 eV, which is lower than the 6.78 eV from the direct dissociation pathway.
Overall, our results show that targeted fine-tuning enables an MLIP to identify plausible intermediates and transition-state structures and to efficiently screen defect diffusion pathways beyond those explicitly included in the training data.
This transferability comes from shared atomic configurations and migration mechanisms between the training data and unexplored pathways.
The MLIP provides a useful tool to extend migration pathway exploration to related defect configurations, especially the formation, migration, and dissociation of defect complexes.

\begin{acknowledgments}
This work was supported by the new faculty start-up grant from UNC-Chapel Hill. 
Computational resources were provided by the Theory and Computation facility of the Center for Functional Nanomaterials (CFN) and the Scientific Data and Computing Center at Brookhaven National Laboratory under contract no. DE-SC0012704, and the Research Computing group from UNC-Chapel Hill.
We also acknowledge the computational resources provided by NCShare, which is supported by National Science Foundation (NSF) grants OAC-2201525, OAC-2201105, and OAC-2430141.
\end{acknowledgments}

\nocite{*}

\bibliography{diamond_defect_diffusion}% Produces the bibliography via BibTeX.

\end{document}